\documentclass{article}

\usepackage{arxiv}

\usepackage[utf8]{inputenc} 
\usepackage[T1]{fontenc}    
\usepackage{hyperref}       
\usepackage{url}            
\usepackage{booktabs}       
\usepackage{amsfonts}       
\usepackage{nicefrac}       
\usepackage{microtype}      
\usepackage{lipsum}		
\usepackage{graphicx}
\usepackage{natbib}
\usepackage{doi}
\usepackage{amsmath,amsfonts}

\title{Perspectives and Challenges of Scaled Boolean Spintronic Circuits Based on Magnetic Tunnel Junction Transducers}

\author{ \href{https://orcid.org/0000-0002-6389-2689}{\includegraphics[scale=0.06]{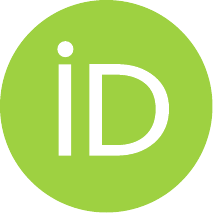}\hspace{1mm}Fanfan Meng} \\
	Imec, Leuven, Belgium\\
 KU Leuven, Leuven, Belgium\\
	\texttt{fanfan.meng@imec.be} \\
	\And
	Siang-Yun Lee \\
	EPFL, Lausanne, Switzerland
	\AND
	Odysseas Zografos \\
	Imec, Leuven, Belgium \\
	\And
	Mohit Gupta \\
	Imec, Leuven, Belgium \\
	 \And
	Van D. Nguyen \\
	Imec, Leuven, Belgium \\
 \And
	Giovanni De Micheli \\
	 EPFL, Lausanne, Switzerland \\
  \And
	Sorin Cotofana \\
	TU Delft, Delft, The Netherlands \\
   \And
	Inge Asselberghs \\
	Imec, Leuven, Belgium \\
 \And
	Christoph Adelmann \\
	Imec, Leuven, Belgium \\
 \And
	Gouri Sankar Kar \\
	Imec, Leuven, Belgium \\
 \And
	Sebastien Couet \\
	Imec, Leuven, Belgium \\
 \And
 \href{https://orcid.org/0000-0002-7088-2075}{\includegraphics[scale=0.06]{orcid.pdf}\hspace{1mm}Florin Ciubotaru}\\
	Imec, Leuven, Belgium \\
    \texttt{Florin.Ciubotaru@imec.be} \\
}

\renewcommand{\shorttitle}{}

\hypersetup{
pdftitle={A template for the arxiv style},
pdfsubject={q-bio.NC, q-bio.QM},
pdfauthor={David S.~Hippocampus, Elias D.~Striatum},
pdfkeywords={First keyword, Second keyword, More},
}

\begin{document}
\maketitle

\begin{abstract}
	This paper addresses the following question: Can  spintronic circuits based on Magnetic Tunnel Junction (MTJ) transducers outperform their state-of-the-art CMOS counterparts? To this end, we use the EPFL combinational benchmark sets, synthesize them in 7 nm CMOS and in MTJ transducer based spintronic technologies, and compare the two implementation methods in terms of Energy-Delay-Product (EDP). To fully utilize the technologies’ potential, CMOS and spintronic implementations are built upon standard Boolean and Majority Gates, respectively. For the spintronic circuits, we assumed that domain conversion (electric/magnetic to magnetic/electric) is performed by means of MTJs and the computation is accomplished by domain wall based majority gates, and considered two EDP estimation scenarios: (i) Uniform Benchmarking, which ignores the circuit's internal structure and only includes domain transducers’ power and delay contributions into the calculations, and (ii) Majority-Inverter-Graph Benchmarking, which also embeds the circuit structure, the associated critical path delay and energy consumption by DW propagation. Our results indicate that for the uniform case, the spintronic route is better suited for the implementation of complex circuits with few inputs and outputs. 
On the other hand, when the circuit structure is also considered via majority and inverter synthesis, our analysis clearly indicates that in order to match and eventually outperform CMOS performance, MTJ  transducers' efficiency has to be improved by 3-4 orders of magnitude. While it is clear that for the time being the MTJ-based-spintronic way cannot compete with CMOS, further technological transducer developments may tip the balance, which, when combined with information non-volatility, may make spintronic implementation for certain applications that require a large number of calculations and have a rather limited amount of interaction with the environment.  
\end{abstract}


\section{Introduction}
The sharp increase in electronic equipment used daily across the globe, from end-user devices to data centers, and the associated energy consumption has led to a craving for more energy-efficient computing devices\cite{Belkhir}. However, the current Moore’s law epitomized miniaturization of microelectronic circuits that rely on CMOS transistors has been gradually limited due to increasing power density and associated chip heating\cite{Gelsinger}. Therefore, intensive research has been devoted to exploring alternative devices \cite{Nikonov.2015,Chen} such as 2D material channel FETs\cite{Radisavljevic}, Mott FET\cite{Ahn}, excitonic device\cite{Dorow}, \textit{etc}. Spintronic devices centered on nanomagnets are seen as a promising category of beyond CMOS devices for (1) the ultra-low energy associated with magnetization dynamics and nanomagnet switching; (2) high endurance; (3) non-volatility to counteract leakage power; (4) capability to build more expressive logic gates (\textit{e.g}., majority gates); and (5) applicability to both traditional and emerging architectures\cite{Dieny}. In the past decade, numerous spintronic logic concepts have been proposed and demonstrated for realizing Boolean logic gates, utilizing, \textit{e.g}., dipolar interactions between nanomagnets, interactions between domain walls, interference of spin waves, and Magneto-Electric Spin-Orbit (MESO) logic\cite{Sivasubramani,meso,zhaochu,Dieny,Barman.2021}.

\begin{figure}[b]
\includegraphics[width=0.9\textwidth]{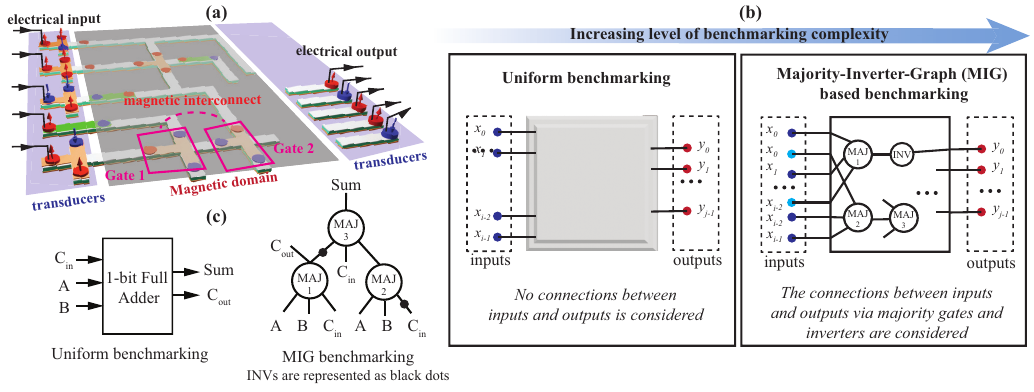}
\centering\caption{(a) A schematic of a hybrid-CMOS-spintronic logic circuit: charged-based information is first converted to magnetic information carriers (\textit{e.g}., domain wall, spin waves, magnetization) via transducers. Then, the computation is achieved by information  carriers’ interaction within the magnetic domain, and finally, the resultant magnetic information is converted back to electrical outputs via transducers. (b) Two benchmarking approaches with different levels of circuit abstractions. (c) Full Adder representation in Uniform and MIG benchmarking, respectively.}
\end{figure}

\indent However, for the time being, there is no concept for a full spintronic computer, which incorporates logic, memory, and interconnects using exclusively magnetic signals\cite{Dieny}. Therefore, it is envisaged that spintronic logic devices will be utilized in hybrid CMOS-spintronic systems where signal interconversion between magnetic and electrical domains via transducers takes place as illustrated in Fig. 1a. The performance of such hybrid systems, in terms of energy consumption and computing throughput, will highly depend on the utilized conversion mechanisms and the number of interconversions needed to perform the computation. Although many spintronic concepts have been proven to materialize in individual logic gates, their integration into CMOS systems, \textit{i.e}., the development of corresponding transducers, is at various stages of maturity.   Up to date, Magnetic Tunnel Junctions (MTJs) that are the key elements in Magnetic Random-Access Memory (MRAM), are the only transducers demonstrated in fully integrated, scaled, and CMOS-compatible Domain Wall (DW) based spintronic logic devices\cite{Raymenants.2020}. Hence, in this work, which attempts to evaluate the targets and challenges of building efficient spintronic Boolean logic circuits from the transducer perspective, we make use of MTJs as a discussion vehicle. Specifically, the Energy-Delay-Product (EDP) is used as a figure of merit to compare a collection of spintronic logic circuits using MTJs as input/output transducers to 7 nm node CMOS technology. As depicted in Fig. 1b, different levels of circuit abstractions are applied to the spintronic circuits, namely Uniform and Majority-Inverter-Graph (MIG)-based benchmarking, to gain insights on the energy-delay cost contributions from different sources.

\section{Overview of the Benchmarking Strategies}
The benchmarking evaluations are carried out in order of increasing complexity; as the analysis progresses, more contributors to the total EDP of the spintronic circuits are considered. We start with the uniform benchmarking (Fig. 1b), which: (i) considers only the energy and delay associated with input ($x_i$) and output ($y_j$) transducers, \textit{i.e}., the switching and detection of the magnetization orientation of MTJs’ free layers and (ii) disregards the magnetic circuits between inputs and outputs, including intermediate spin logic gates and magnetic interconnects. Additionally, this method takes into account the minimum number of transducers required in a hybrid CMOS-spintronic circuit, as defined by the circuit’s function. As an example, a Full Adder (FA) (Fig. 1c) adds together two binary digits plus a carry-in digit to produce a sum and carry-out digit and therefore requires at least three inputs and two outputs. By only considering the minimum number of transducers required in the system, the uniform method provides a system EDP lower bound as well as a minimum target for transducer efficiency for which spin-hybrid circuits can outperform CMOS. 
The minimum transducer efficiency target derived with this method holds true regardless of the paradigm (\textit{e.g}., spin wave computing, plasmonic computing, MESO logic) and different circuit implementations, and hence the method is known as uniform benchmarking\cite{nikonov}.

To get better EDP estimates, we need to further consider the actual structure of the benchmark circuits. Given that spintronics provides natural support for inverter\cite{zhaochu} and majority (MAJ) gate\cite{stmg} implementations, which together form a universal gate set, we make use of such elements to describe the internal organization of the circuit. MAJ gate operates according to the majority voting principle, can emulate both logic AND and OR operations, and promises circuits with higher computational density\cite{Zografos.2015}. Thus, to fully exploit the gains brought by majority functions, instead of using standard logic synthesis tools based on AND, OR, XOR, and NAND gates, we employ a customized logic synthesis tool known as an Majority-Inverter-Graph (MIG), which provides guidelines on the realization of logic circuits using majority gates and inverters\cite{MIG}. 
MIGs provide the number of gates required and how they are connected at the logic level, however, they do not reveal the physical placements and routings of these gates. 
Again, using the FA as an example, the additional cost related to four repeated inputs, three MAJs and two INVs is considered (Fig. 1c). We further divide this benchmarking into two phases. In the first one, we consider the additional energy cost due to repeated inputs, and the delay in the magnetic domain. In the second phase, we additionally include the energy cost related to information propagation in the magnetic domain. Details on the made assumptions are provided in section IV. 

\begin{figure*}
\includegraphics[width=0.88\textwidth]{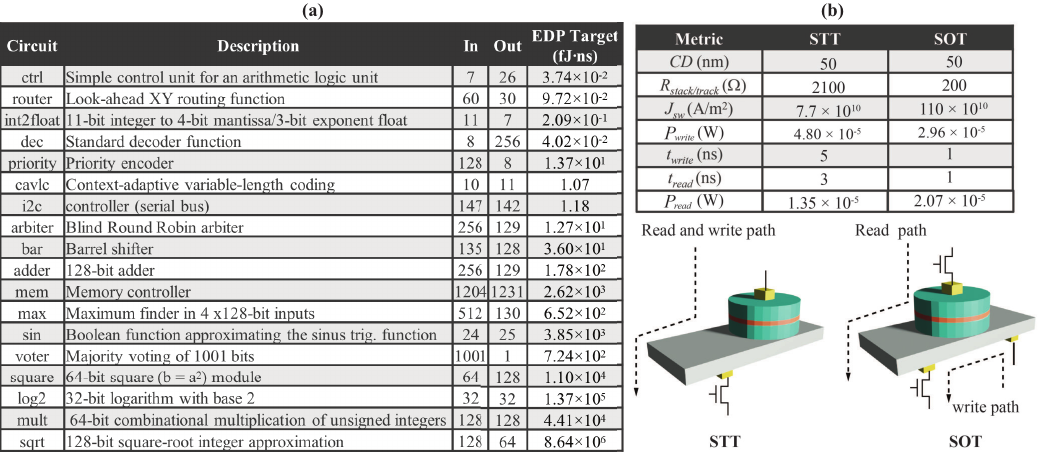}
\centering\caption{(a) List of circuits with their descriptions, number of inputs and outputs, and the Energy-Delay-Product (EDP) from CMOS synthesis. The EDP of CMOS 7 nm node technology sets the target for spintronic circuits. (b) Specifications and schematics of STT- and SOT-MTJs\cite{couet2021beol,sakhare2018enablement,mohit}. The thickness of the SOT track is 7 nm.  }
\end{figure*}

In both benchmarking studies, we use a collection of representative combinational logic circuits from the EPFL Combinational Benchmarking Suite (Fig. 2a) with a large variation in sizes, complexity levels, and input/output (I/O) ratios\cite{Amaru15benchmark}. These circuits were first synthesized with commercial software in 7nm CMOS technology to provide a comparison base of spintronic circuits with CMOS counterparts implemented in leading-edge mass production technology\cite{raghavan2015holisitic}. The CMOS synthesis is optimized for low-power operation and the EDP is used to set targets for their spintronic counterparts. As for spintronic circuit transducers, we considered state-of-art Spin-Transfer-Torque (STT)\cite{sakhare2018enablement}- and Spin-Orbit-Torque (SOT)\cite{couet2021beol}-based MTJ technologies. The energy and delay of individual MTJ’s writing and reading\cite{mohit} are summarized in Fig. 2b. 

\section{Uniform Benchmarking}
In uniform benchmarking, where only the energy and delay cost of the minimum number of transducers are considered, the spintronic circuits' EDP is defined as
\begin{equation*}
   EDP_{spin}=\underbrace{(n_{in}\times E_{w}+n_{out}\times E_{r})}_{total\:energy}\times
  \underbrace{ (t_{w}+t_{r})}_{total\:delay},
\end{equation*}
where $n_{in}$ and $n_{out}$ are the number of inputs and outputs as listed in Fig. 2a and $E_{w}$, $E_{r}$, $t_{w}$, $t_{r}$ are the energy and delay associated with the writing and reading operations on an individual MTJ. 
Fig. 3a depicts EDP values calculated for CMOS, SOT- and STT-MTJ-enabled spintronic circuits. 
The circuits highlighted with a red background, such as ‘ctrl’ and ‘arbiter’ have two to four orders of magnitude higher EDP for spintronic circuits than for CMOS. The EDP cost solely at the transducer interfaces already greatly exceeds the budget set by CMOS and implies that the MTJ performance must be drastically improved for spintronic circuits to match CMOS. However, for circuits highlighted with a green background like ‘log2’ and ‘sqrt’, the spintronic EDP is lower than CMOS allowing a margin for magnetic circuitry to be included  in further MIG benchmarking. 
\begin{figure*}
\includegraphics[width=0.9\textwidth]{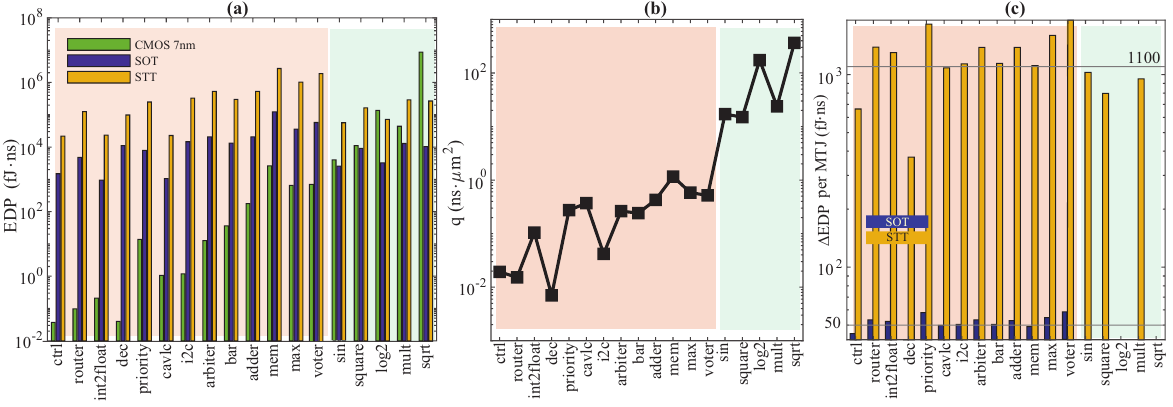}
\centering\caption{Uniform Benchmarking Results. (a) The comparison of EDP between CMOS circuits and SOT- and STT-MTJ mediated spintronic circuits. The circuits with lower or higher EDP compared to CMOS are highlighted in green and red, respectively. (b) $q$ metric is defined to identify potential spintronic circuits that are more efficient than CMOS circuits. (c) The improvement in EDP required from individual MTJs to have the total EDP  comparable to CMOS circuits.   }
\end{figure*}

To identify the circuits with lower/higher EDP compared to CMOS, a metric $q$ is defined as
\begin{equation}
 q=\frac{(area \times delay)_{\mathrm{cmos}}}{(n_{in}+n_{out})}.
\end{equation}
Fig. 3b indicates that circuits driven by SOT-MTJs with q$>$10 have a lower EDP than CMOS, therefore could be considered more advantageous with spin logic. Since the area-delay product of CMOS circuits is a measure of circuits’ complexity, a larger $q$ implies that a spintronic avenue is more appropriate for the implementation of complicated circuits with few inputs/outputs, thus requiring less transduction. Note that in this approach, the spintronic systems do not include the logic circuit itself, whereas this is included for CMOS implementations. For complex circuits, this leads to a larger advantage of spintronic circuits, which is expected to reduce when the logic circuit is considered as demonstrated in the MIG benchmarking section. For circuits where the energy-delay cost in I/O transduction already exceeds the EDP budget set by CMOS, we apportion the difference in EDP to each individual MTJ,
\begin{equation}
    \Delta EDP_{per \, MTJ} =\frac{EDP_{spin}-EDP_{cmos}}{n_{in}+n_{out}}
\end{equation}
leading to EDP performance upper bound  of individual SOT/STT MTJs. As plotted in Fig. 3c, an average decrease of 50$\times$ (SOT) to 1100$\times$ (STT) is required in terms of EDP for single MTJ devices. Note that reducing writing and reading delays will have a much stronger impact on the EDP when compared to improving power consumption since a longer delay also increases energy consumption.
Regarding energy consumption at transduction interfaces only (as presented in Fig. 4), in SOT-MTJ-driven spintronic circuits, on average 63\% of the energy is consumed by the input transducers, while in STT-MTJ-driven circuits, 84 \% of the energy is consumed at the input interfaces.

\begin{figure}[!t]
\centering
\includegraphics[width=2.5in]{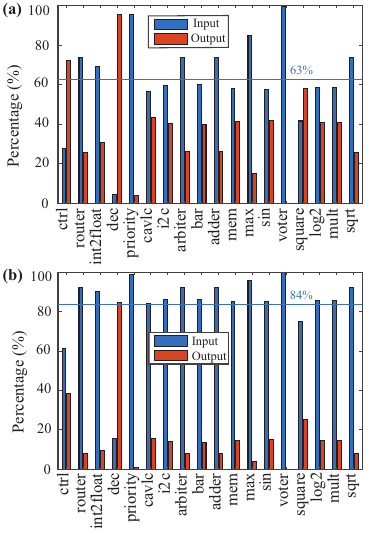}
\caption{Percentage energy consumption by inputs and outputs transducers using (a) SOT- (b) STT- MTJs.}
\label{fig1}
\end{figure}

\section{Majority-Inverter-Graph (MIG) Benchmarking}
In addition to the minimum number of I/O transducers considered in uniform benchmarking, we bring into the picture  the internal spintronic circuit structure and the associated energy consumption and delay overheads by means of MIG-based synthesis. 
All benchmark implementations are optimized to minimize the number of MAJ gates, and we assumed that MAJ and INV have infinite fan-out and cascading capability in the magnetic domain\cite{MIG}. Note this is a very optimistic assumption for spintronic logic gates, as currently there is no experimental demonstration of these capabilities. Fig. 5a and 5b display a section and the full MIG of the ‘ctrl’ circuit, respectively, which is one of the smallest circuits in the benchmark set. Primary inputs ($x_i$), majority gates ($n_j$), outputs ($y_k$), and inverters are depicted as blue squares, black dots, red squares, and blue lines, respectively. Other assumptions used in this benchmarking will be explained using this circuit as an example. 

\begin{figure}[!t]
\centering
\includegraphics[width=2.5in]{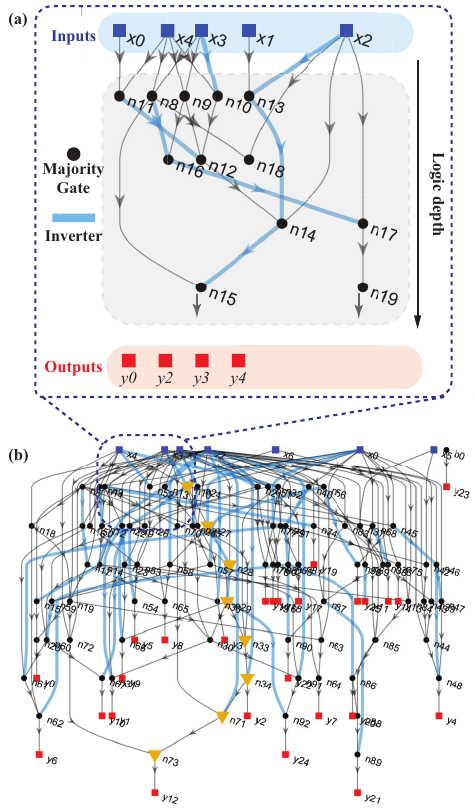}
\caption{Majority Inverter Graph (MIG). (a) A small section of the MIG graph from ‘ctrl’ circuit. Inputs ($x_i$), majority gates ($n_j$), inverters, and outputs ($y_k$) are shown as blue squares, black dots, blue lines, and red squares, respectively. (b) The full MIG graph of the ‘ctrl’ circuit. The longest path is marked by yellow triangles.}
\end{figure}
\begin{figure}[!t]
\centering
\includegraphics[width=2.3in]{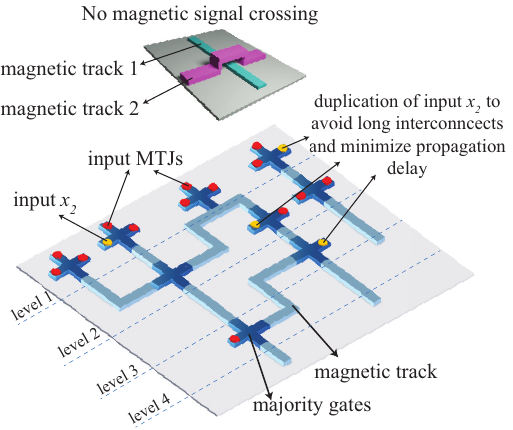}
\caption{Schematics illustrating the assumption of input transducer duplication.  }
\label{fig1}
\end{figure}

\subsection{Duplication of inputs and delay in the magnetic domain}
First, as shown in Fig. 5a, each independent input ($x_i$) can potentially drive multiple gates at different logic depths which are defined as the maximum number of gates a signal needs to travel from the primary inputs to the destination. For instance, primary input $x_2$ will drive majority gates $n_{13}$, $n_{18}$ and $n_{17}$ that are  at different logic depths. 
As illustrated in Fig. 6, since 3D magnetic signal crossing is not available, to supply the primary inputs to deeper-level gates, long magnetic interconnects are needed to bypass gates at shallower depths. To minimize the delay due to signal propagation in long magnetic interconnects, we assume a duplication of each primary input at the place it is needed. The resultant number of input transducers required is summarized in Fig. 7a. Compared to the number of inputs considered in the uniform benchmarking, an average factor of 10$\times$ is found for the analyzed circuits, which  leads to a similar increase in the EDP. As shown in Fig. 7b and 7c, duplication of inputs suggested by MIG synthesis also results in the fact that, at the transducing interfaces, more than 90\% of the energy is spent at the input stage for both SOT- and STT-MTJ-driven spintronic circuits, \textit{i.e}., improving the energy performance of the input transducers will have a larger impact on the overall EDP performance.
\begin{figure*}
\includegraphics[width=0.85\textwidth]{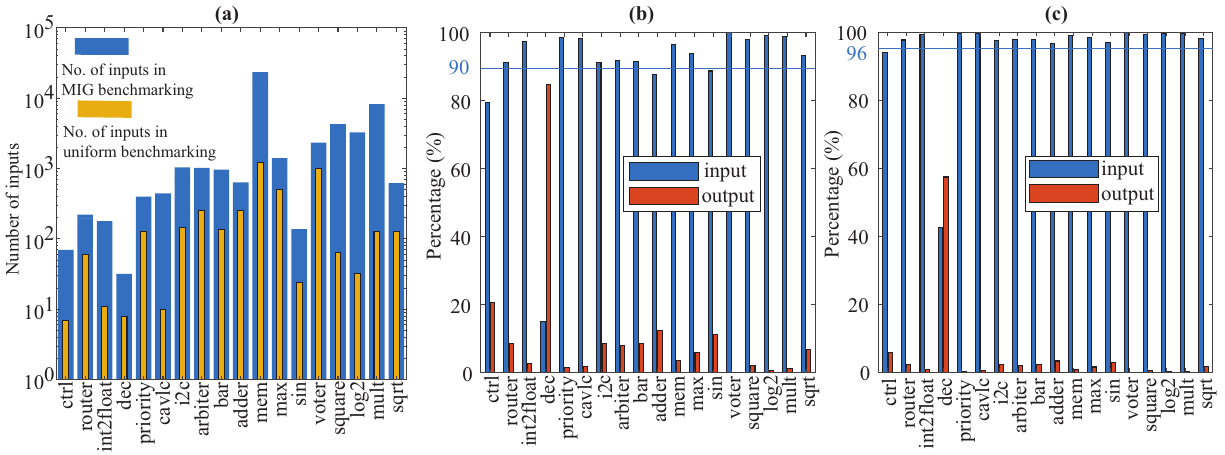}
\centering\caption{ (a) Comparison of number of inputs considered in uniform and MIG-based benchmarking. (b)-(c) Percentage energy consumption by inputs and outputs at transducing interfaces for spintronic circuits using SOT- and STT- MTJs, respectively. }
\end{figure*}
\begin{figure*}
\includegraphics[width=0.85\textwidth]{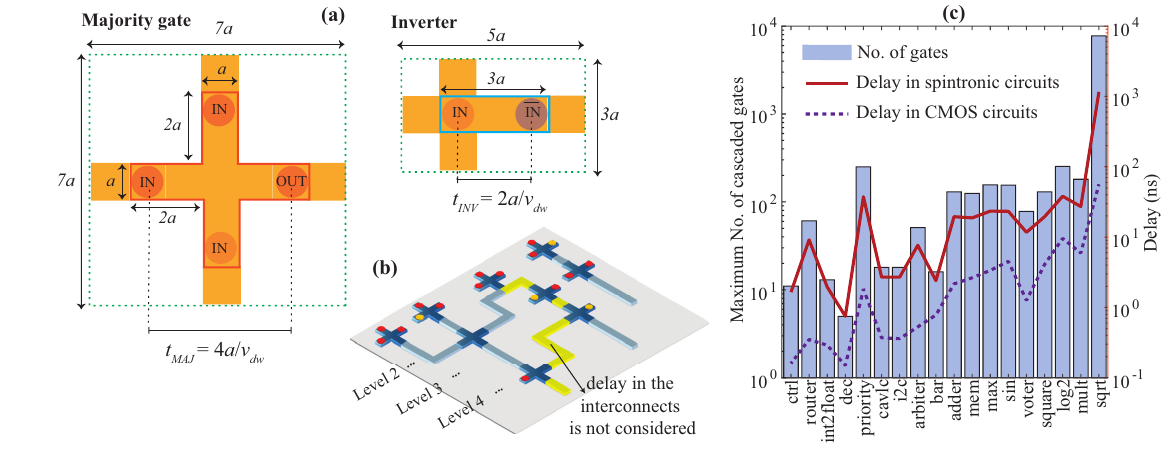}
\centering\caption{(a) Delay and area assumptions for majority gates and inverters based on domain wall logic driven by STT-MTJ transducers. (b) A schematic illustrating the assumption that delay in the interconnects is not considered. (c) The maximum number of gates cascaded between inputs and outputs in different circuits. The delay in the CMOS circuits and the delay in the spintronic circuits without considering the DW propagation time in the interconnects.}
\end{figure*}

Second, the delay of the circuits is estimated by determining the maximum logic depth between inputs and outputs. For example, the longest path for the complete 'ctrl' circuit (marked by yellow triangles in Fig. 5b) is formed by 11 gates (8 MAJs and 3 INVs). Adopting the most common geometries proposed for a domain wall based spin torque majority gate and an inverter, as graphically depicted in Fig. 8a\cite{stmg, zhaochu}, their delays are estimated to be $t_{maj}=4a/v_{dw}$ and $t_{inv}=2a/v_{dw}$, where $a$ is the critical dimension of MTJs and $v_{dw}$ is the domain wall velocity. In this benchmarking, we assume $a=50$ nm, which is the most common reported critical dimension for STT-MTJ\cite{sakhare2018enablement} and $v_{dw}=$750 m/s \cite{dw_velocity,RacetrackMemory}, a typical value for domain wall velocity reported in materials that are compatible with MTJ structures. Assuming these gates can be cascaded together directly without additional interconnects that add to the delay (Fig. 8b), the total computation time in the magnetic domain is estimated for the 'ctrl' circuit to be $t_{mag}=8\times t_{maj}+3\times t_{inv}$. The longest path for each circuit, \textit{i.e}., the maximum number of cascaded gates, are presented as blue bars in Fig. 8c. The corresponding delay in the magnetic domain calculated using these assumptions and the total delay in CMOS circuits are plotted as lines in Fig. 8c as well.  The data shows that the computing time in spintronic circuits that has a large impact on the overall EDP is already one order of magnitude greater than the total delay seen in CMOS circuits even without considering the potential propagation time in the interconnects between gates. As mentioned earlier, these interconnect can be very long due to the lack of 3D magnetic signal crossing. 

Taking the two additional components in magnetic circuitry revealed by MIG, \textit{i.e.}, the increase in the number of input transducers and the delay in the magnetic domain, the EDP of spintronic circuits becomes
\begin{equation*}
      EDP_{spin}= \\ \underbrace{(n_{mig\_in}\times E_{w}+n_{out}\times E_{r})}_{total\:energy}\times 
  \underbrace{ (t_{w}+t_{r}+t_{mag})}_{total\:delay},   
\end{equation*}
 where $n_{mig\_in}$  is the number of input transducers required by the MIG synthesis and $t_{mag}=d_{maj}\times t_{maj}+d_{inv}\times t_{inv}$ is the total operation time in the magnetic domain. $d_{maj}$ and $d_{inv}$ are the number of majority gates and inverters on the critical path of each circuit. Note, at this stage, we assumed that propagating domain walls in the magnetic domain requires no energy, which is relevant to the logic concept based on exchange-driven domain wall automation\cite{exchangedw}.
 Fig. 9a presents the EDP of both CMOS circuits and spintronic circuits. Now, the EDPs for all investigated circuits in spintronics are on average two orders of magnitude higher. As previously, we evenly distribute the EDP excess over the budget set by CMOS to all MTJs. As a result, EDP performances of SOT- and STT-MTJs need to be reduced by about 790$\times$ and  8700$\times$, respectively (see Fig. 9b).
 
 Additionally, we calculate the area of the spintronic circuit by considering only the footprints of majority and inverter gates, whereas the area related to interconnects is neglected, thus since the MIG synthesis targeted gate count minimization, we calculate the area lower bound.  The area of individual MAJ and INV is is estimated as 49$a^2$ and 15$a^2$\cite{stmg}, respectively as indicated in Fig. 8a. 
 \begin{figure*}
\includegraphics[width=1\textwidth]{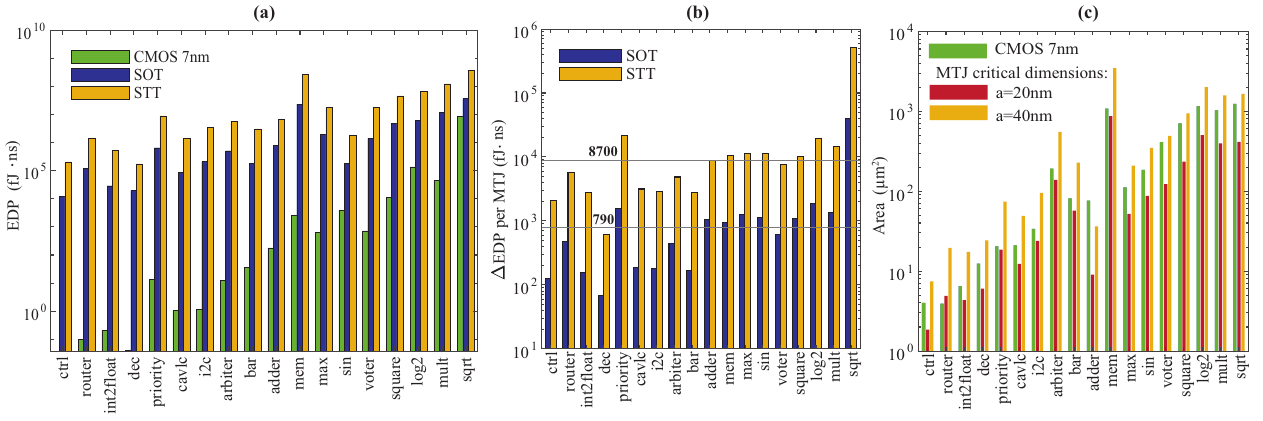}
\centering\caption{MIG benchmarking results considering the increased number of inputs and delay in the magnetic domain. (a) The comparison of EDP between CMOS circuits and SOT- and STT-MTJ mediated spintronic circuits. (b) The improvement in EDP required from individual MTJs to have the total EDP  comparable to CMOS circuits.  (c) Area comparison for CMOS and spintronic circuits. MTJ critical dimensions of 20 nm and 40 nm are assumed for the area estimation for spintronic circuits.}
\end{figure*}
 In Fig. 9c, we compare the area of CMOS and spintronic circuits for MTJ critical dimensions of 40 and 20 nm. 
 The results indicate that even without considering the real physical layout of the circuits, the MTJ critical dimensions have to be at most 20 nm to surpass CMOS circuits in terms of area compactness. Note that area calculations are done for STT-MTJs transducers as SOT-MTJs-based implementations require an even larger footprint due to their three-terminal design.

\begin{figure*}
\includegraphics[width=1\textwidth]{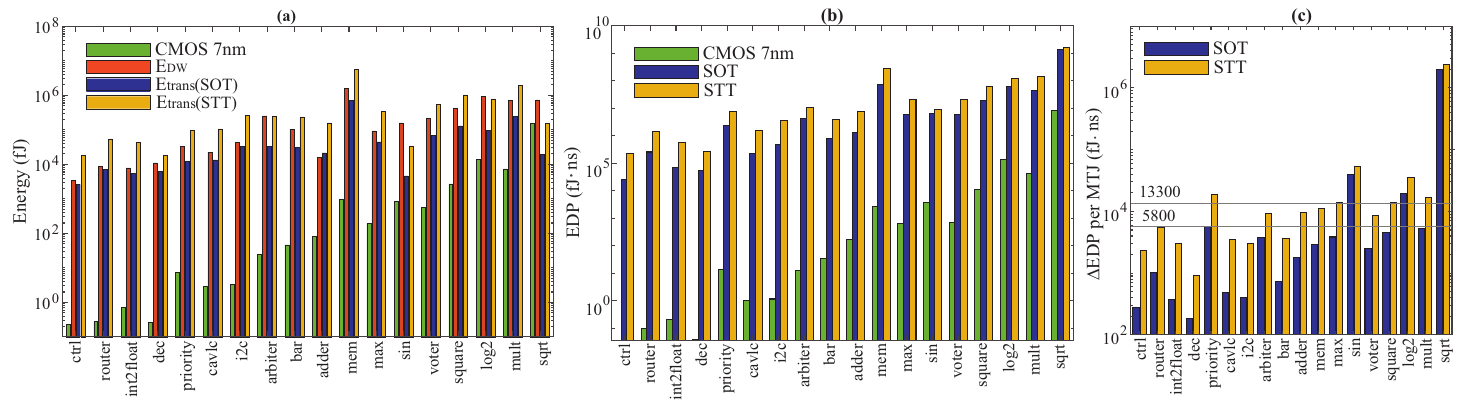}
\centering\caption{(a) Comparisons of energy consumption in the CMOS circuits with the energy consumed in the magnetic domain and the transducing interfaces in the SOT-MTJs and STT-MTJs driven spintronic circuits, respectively. (c) EDP for CMOS circuits and spintronic circuits including the energy consumption in the magnetic domain. (d) The improvement in EDP required from individual MTJs to have the total EDP  comparable to CMOS circuits.}
\end{figure*}

\subsection{The energy required to propagate domain walls}
Finally, we consider the energy required to propagate domain walls in the logic gates. In 2020, Luo \textit{et al}., have demonstrated SOT-current driven domain wall logic, and here we adopt the same method to calculate the energy consumption per operation of the gate, which is the power-delay product of the current in the bottom Pt layer\cite{zhaochu}. The energy required to push domain walls across one arm of the majority gate is
\begin{equation*}
    E_{arm}=\frac{\rho J^2 w h L^2}{v_{dw}}
\end{equation*}
where $\rho$ = 30 $\mu\Omega$cm is the resistivity of the Pt layer\cite{zhaochu},  $v_{dw}=$750 m/s is the domain wall velocity, and $J$ = 3$\times{10}^{12}$ A/m$^2$ is the current density required to achieve this domain wall velocity\cite{RacetrackMemory,dw_velocity}.  We assume that the length and the width of the domain wall track are $w=a$ and $L=2a$ (see Fig. 8a), and $h$ = 5 nm is the thickness of the Pt layer \cite{zhaochu}. It is worth noting that for a cascaded network to work, the network paths have to be resistively balanced (\textit{e.g}., by making use of clipping resistors\cite{Vaysset}) so that the same current can flow through all devices in the network. Here, we only consider the minimum current density required to push domain walls; the additional energy cost of clipping resistance is not considered. Hence, the energy required by an individual majority gate and an inverter is $E_{maj}=4E_{arm}$ and $E_{inv}=E_{arm}$, respectively. Now the total energy required to drive domain walls in the spintronic circuits is 
\begin{equation*}
          E_{DW}= n_{maj}\times E_{maj} + n_{inv}\times E_{inv}
\end{equation*}
where $n_{maj}$ and $n_{inv}$ are the total number of majority gates and inverters needed to build the circuit. 
In Fig. 10a, for each spintronic circuit, the energy consumption to push domain walls  in the logic gates ($E_{DW}$) and the energy cost of SOT or STT-MTJs transducers($E_{trans}$),  are compared with the total energy consumption of the corresponding CMOS circuit. 
The energy consumption within the magnetic domain is of the same order of magnitude as the energy spent at the transducing interfaces, which leads to a further 2$\times$ EDP increase (Fig. 9b). Again, we evenly distribute the EDP excess over the budget set by CMOS to all MTJs and, as a result, the SOT- and STT-MTJs performance needs to be improved by  5800$\times$, and  13300$\times$, respectively (see Fig. 9c).

\section{Conclusion}
In this work, we  evaluate the challenges and targets of building Boolean spintronic circuits from the perspective of transducers and specifically focused on MTJs, the only scalable option up to date. By only considering components revealed at the MIG logic synthesis level, the EDP performance of SOT- and STT-MTJs needs to be reduced by $\sim$ 5800$\times$, and $\sim$ 13300$\times$, respectively, and the critical dimension of MTJs needs to be reduced to 20 nm to be more compact than CMOS circuits. It is also important to note that there are still major contributors to the EDP yet to be considered. First, at the logic synthesis level, we have yet to consider the fan-out and cascading limitations of spintronic logic gates, which will lead to duplications of sections of circuits and hence an increase in the number of input transducers and logic gates required. Second, on the physical layout level, a main limitation for some of the spintronic concepts, \textit{e.g}., domain walls and spin waves, is the lack of information signal crossing in magnets at the nanoscale. To layout such circuits without any line crossover, duplication of circuits and long interconnects are expected, which will add significant delay and energy costs. In our benchmarking, no energy nor delay related to the interconnects is considered.

In conclusion, a synergy of effort from various aspects is required to build efficient spintronic circuits. First, efficient transducers are vital in the construction of spintronic circuits. The performance of MTJs must be strongly enhanced to be considered viable choices in the traditional Boolean logic architecture. Voltage-based transducers may be able to bridge this gap\cite{meso,prasad}. As important as transducers are, the delay in the magnetic domain must be drastically improved. In addition to the enhancement of the speed of information carriers, such as increasing the domain wall velocities, efforts need to be put into minimizing the interconnect length, which will require the abilities of fan-out, cascading, and signal crossing of magnetic information carriers, which are not well addressed in the current literature. Wave-pipelining is one option for increasing the throughput of spintronic circuits\cite{pipe}, however, it requires uniform propagation delay between any gates from two logic depths, \textit{i.e}., the same interconnect length, which demands more stringent design flexibility of interconnects. Spintronic concepts such as MESO\cite{meso} that only require charge interconnects are preferred. In the magnetic domain, the energy required to propagate domain walls is also a significant contributor to overall energy consumption. Fundamental studies on reducing the current density while maintaining high speed as well as new mechanisms to propagate magnetic information are in demand. More computing architecture explorations that can exploit the non-volatility and stochasticity of spintronic devices, in the meantime require a large number of calculations and a limited amount of interaction with the environment may compare more favorably with CMOS circuits. 

\section{Acknowledgement}
This work was supported by imec's Industrial Affiliation Program on Exploratory Logic Devices. It has also received funding from the European Union’s Horizon Europe research and innovation programme within the project SPIDER under grant agreement No 101070417.

\bibliographystyle{unsrtnat}
\bibliography{references}  

\begin{thebibliography}{28}
\providecommand{\natexlab}[1]{#1}
\providecommand{\url}[1]{\texttt{#1}}
\expandafter\ifx\csname urlstyle\endcsname\relax
  \providecommand{\doi}[1]{doi: #1}\else
  \providecommand{\doi}{doi: \begingroup \urlstyle{rm}\Url}\fi

\bibitem[Belkhir and Elmeligi(2018)]{Belkhir}
Lotfi Belkhir and Ahmed Elmeligi.
\newblock {Assessing ICT global emissions footprint: Trends to 2040 \&
  recommendations}.
\newblock \emph{Journal of Cleaner Production}, 177:\penalty0 448--463, 2018.
\newblock ISSN 0959-6526.
\newblock \doi{10.1016/j.jclepro.2017.12.239}.

\bibitem[Gelsinger(2001)]{Gelsinger}
Patrick~P. Gelsinger.
\newblock {Microprocessors for the New Millennium: Challenges, Opportunities,
  and New Frontiers}.
\newblock \emph{2001 IEEE International Solid-State Circuits Conference. Digest
  of Technical Papers. ISSCC (Cat. No.01CH37177)}, pages 22--25, 2001.
\newblock \doi{10.1109/isscc.2001.912412}.

\bibitem[Nikonov and Young(2015)]{Nikonov.2015}
Dmitri~E. Nikonov and Ian~A. Young.
\newblock {Benchmarking of Beyond-CMOS Exploratory Devices for Logic Integrated
  Circuits}.
\newblock \emph{IEEE Journal on Exploratory Solid-State Computational Devices
  and Circuits}, 1:\penalty0 3--11, 2015.
\newblock \doi{10.1109/jxcdc.2015.2418033}.

\bibitem[Chen(2022)]{Chen}
An~Chen.
\newblock {Beyond-CMOS roadmap—from Boolean logic to neuro-inspired
  computing}.
\newblock \emph{Japanese Journal of Applied Physics}, 61\penalty0
  (SM):\penalty0 SM1003, 2022.
\newblock ISSN 0021-4922.
\newblock \doi{10.35848/1347-4065/ac5d86}.

\bibitem[Radisavljevic et~al.(2011)Radisavljevic, Radenovic, Brivio,
  Giacometti, and Kis]{Radisavljevic}
B.~Radisavljevic, A.~Radenovic, J.~Brivio, V.~Giacometti, and A.~Kis.
\newblock {Single-layer MoS2 transistors}.
\newblock \emph{Nature Nanotechnology}, 6\penalty0 (3):\penalty0 147--150,
  2011.
\newblock ISSN 1748-3387.
\newblock \doi{10.1038/nnano.2010.279}.

\bibitem[Ahn et~al.(2003)Ahn, Triscone, and Mannhart]{Ahn}
C.~H. Ahn, J.-M. Triscone, and J.~Mannhart.
\newblock {Electric field effect in correlated oxide systems}.
\newblock \emph{Nature}, 424\penalty0 (6952):\penalty0 1015--1018, 2003.
\newblock ISSN 0028-0836.
\newblock \doi{10.1038/nature01878}.

\bibitem[Dorow et~al.(2018)Dorow, Leonard, Fogler, Butov, West, and
  Pfeiffer]{Dorow}
C.~J. Dorow, J.~R. Leonard, M.~M. Fogler, L.~V. Butov, K.~W. West, and L.~N.
  Pfeiffer.
\newblock {Split-gate device for indirect excitons}.
\newblock \emph{Applied Physics Letters}, 112\penalty0 (18):\penalty0 183501,
  2018.
\newblock ISSN 0003-6951.
\newblock \doi{10.1063/1.5021488}.

\bibitem[Dieny et~al.(2020)Dieny, Prejbeanu, Garello, Gambardella, Freitas,
  Lehndorff, Raberg, Ebels, Demokritov, Akerman, Deac, Pirro, Adelmann, Anane,
  Chumak, Hirohata, Mangin, Valenzuela, Onbaşlı, d’Aquino, Prenat,
  Finocchio, Lopez-Diaz, Chantrell, Chubykalo-Fesenko, and Bortolotti]{Dieny}
B.~Dieny, I.~L. Prejbeanu, K.~Garello, P.~Gambardella, P.~Freitas,
  R.~Lehndorff, W.~Raberg, U.~Ebels, S.~O. Demokritov, J.~Akerman, A.~Deac,
  P.~Pirro, C.~Adelmann, A.~Anane, A.~V. Chumak, A.~Hirohata, S.~Mangin,
  Sergio~O. Valenzuela, M.~Cengiz Onbaşlı, M.~d’Aquino, G.~Prenat,
  G.~Finocchio, L.~Lopez-Diaz, R.~Chantrell, O.~Chubykalo-Fesenko, and
  P.~Bortolotti.
\newblock {Opportunities and challenges for spintronics in the microelectronics
  industry}.
\newblock \emph{Nature Electronics}, 3\penalty0 (8):\penalty0 446--459, 2020.
\newblock \doi{10.1038/s41928-020-0461-5}.

\bibitem[Sivasubramani et~al.(2019)Sivasubramani, Mattela, Pal, and
  Acharyya]{Sivasubramani}
Santhosh Sivasubramani, Venkat Mattela, Chandrajit Pal, and Amit Acharyya.
\newblock {Nanomagnetic logic design approach for area and speed efficient
  adder using ferromagnetically coupled fixed input majority gate}.
\newblock \emph{Nanotechnology}, 30\penalty0 (37):\penalty0 37LT02, 2019.
\newblock ISSN 0957-4484.
\newblock \doi{10.1088/1361-6528/ab295a}.

\bibitem[Manipatruni et~al.(2018)Manipatruni, Nikonov, and Young]{meso}
Sasikanth Manipatruni, Dmitri~E Nikonov, and Ian~A Young.
\newblock {Beyond CMOS computing with spin and polarization}.
\newblock \emph{Nature Physics}, 14\penalty0 (4):\penalty0 338--343, 2018.
\newblock ISSN 1745-2473.
\newblock \doi{10.1038/s41567-018-0101-4}.

\bibitem[Luo et~al.(2020)Luo, Hrabec, Dao, Sala, Finizio, Feng, Mayr, Raabe,
  Gambardella, and Heyderman]{zhaochu}
Zhaochu Luo, Aleš Hrabec, Trong~Phuong Dao, Giacomo Sala, Simone Finizio,
  Junxiao Feng, Sina Mayr, Jörg Raabe, Pietro Gambardella, and Laura~J.
  Heyderman.
\newblock {Current-driven magnetic domain-wall logic}.
\newblock \emph{Nature}, 579\penalty0 (7798):\penalty0 214--218, 2020.
\newblock ISSN 0028-0836.
\newblock \doi{10.1038/s41586-020-2061-y}.

\bibitem[Barman et~al.(2021)Barman, Gubbiotti, Ladak, Adeyeye, Krawczyk,
  Gräfe, Adelmann, Cotofana, Naeemi, Vasyuchka, Hillebrands, Nikitov, Yu,
  Grundler, Sadovnikov, Grachev, Sheshukova, Duquesne, Marangolo, Gyorgy,
  Porod, Demidov, Urazhdin, Demokritov, Albisetti, Petti, Bertacco, Schulteiss,
  Kruglyak, Poimanov, Sahoo, Sinha, Yang, Muenzenberg, Moriyama, Mizukami,
  Landeros, Gallardo, Carlotti, Kim, Stamps, Camley, Rana, Otani, Yu, Yu,
  Bauer, Back, Uhrig, Dobrovolskiy, Dijken, Budinska, Qin, Chumak, Khitun,
  Nikonov, Young, Zingsem, and Winklhofer]{Barman.2021}
A~Barman, Gianluca Gubbiotti, Sam Ladak, Adekunle~Olusola Adeyeye, Maciej
  Krawczyk, Joachim Gräfe, Christoph Adelmann, Sorin Cotofana, Azad Naeemi,
  Vitaliy~I Vasyuchka, Burkard Hillebrands, S~A Nikitov, Haiming Yu, Dirk
  Grundler, Alexandr Sadovnikov, Andrew~A Grachev, S~E Sheshukova, Jean-Yves
  Duquesne, Massimiliano Marangolo, Csaba Gyorgy, Wolfgang Porod, V~E Demidov,
  Sergei Urazhdin, Sergej Demokritov, Edoardo Albisetti, Daniela Petti,
  Riccardo Bertacco, Helmut Schulteiss, Volodymyr~V Kruglyak, Vlad~D Poimanov,
  Ashok~Kumar Sahoo, Jaivardhan Sinha, Hyunsoo Yang, Markus Muenzenberg,
  Takahiro Moriyama, Shigemi Mizukami, Pedro Landeros, Rodolfo~Andrés
  Gallardo, Giovanni Carlotti, Joo-Von Kim, Robert~L Stamps, Robert~E Camley,
  Bivas Rana, Y~Otani, Weichao Yu, Tao Yu, Gerrit E~W Bauer, Christian~H Back,
  Goetz~S Uhrig, Oleksandr~V Dobrovolskiy, Sebastiaan~van Dijken, Barbora
  Budinska, Huajun Qin, Andrii Chumak, Aleksandr Khitun, Dmitri~E Nikonov,
  Ian~A Young, Benjamin Zingsem, and Michael Winklhofer.
\newblock {The 2021 Magnonics Roadmap}.
\newblock \emph{Journal of Physics: Condensed Matter}, 2021.
\newblock ISSN 0953-8984.
\newblock \doi{10.1088/1361-648x/abec1a}.

\bibitem[Raymenants et~al.(2020)Raymenants, Wan, Couet, Souriau, Thiam,
  Tsvetanova, Canvel, Asselberghs, Heyns, Nikonov, Young, Pizzini, Nguyen, and
  Radu]{Raymenants.2020}
E.~Raymenants, D.~Wan, S.~Couet, L.~Souriau, A.~Thiam, D.~Tsvetanova,
  Y.~Canvel, I.~Asselberghs, M.~Heyns, D.E. Nikonov, I.A. Young, S.~Pizzini,
  V.D. Nguyen, and I.P. Radu.
\newblock {All-electrical control of scaled spin logic devices based on domain
  wall motion}.
\newblock \emph{2020 IEEE International Electron Devices Meeting (IEDM)},
  00:\penalty0 21.5.1--21.5.4, 2020.
\newblock \doi{10.1109/iedm13553.2020.9372127}.

\bibitem[Nikonov and Young(2012)]{nikonov}
Dmitri~E. Nikonov and Ian~A. Young.
\newblock {Uniform Methodology for Benchmarking Beyond-CMOS Logic Devices}.
\newblock \emph{2012 International Electron Devices Meeting}, pages
  25.4.1--25.4.4, 2012.
\newblock \doi{10.1109/iedm.2012.6479102}.

\bibitem[Nikonov et~al.(2011)Nikonov, Bourianoff, and Ghani]{stmg}
Dmitri~E. Nikonov, George~I. Bourianoff, and Tahir Ghani.
\newblock {Proposal of a Spin Torque Majority Gate Logic}.
\newblock \emph{IEEE Electron Device Letters}, 32\penalty0 (8):\penalty0
  1128--1130, 2011.
\newblock ISSN 0741-3106.
\newblock \doi{10.1109/led.2011.2156379}.

\bibitem[Zografos et~al.(2015)Zografos, Sorée, Vaysset, Cosemans, Amarù,
  Gaillardon, Micheli, Lauwereins, Sayan, Raghavan, Radu, and
  Thean]{Zografos.2015}
Odysseas Zografos, Bart Sorée, Adrien Vaysset, Stefan Cosemans, Luca Amarù,
  Pierre-Emmanuel Gaillardon, Giovanni~De Micheli, Rudy Lauwereins, Safak
  Sayan, Praveen Raghavan, Iuliana~P. Radu, and Aaron Thean.
\newblock {Design and Benchmarking of Hybrid CMOS-Spin Wave Device Circuits
  Compared to 10nm CMOS}.
\newblock \emph{2015 IEEE 15th International Conference on Nanotechnology
  (IEEE-NANO)}, pages 686--689, 2015.
\newblock \doi{10.1109/nano.2015.7388699}.

\bibitem[Amarú et~al.(2016)Amarú, Gaillardon, and Micheli]{MIG}
Luca Amarú, Pierre-Emmanuel Gaillardon, and Giovanni~De Micheli.
\newblock {Majority-based Synthesis for Nanotechnologies}.
\newblock \emph{2016 21st Asia and South Pacific Design Automation Conference
  (ASP-DAC)}, pages 499--502, 2016.
\newblock \doi{10.1109/aspdac.2016.7428061}.

\bibitem[Couet et~al.(2021)Couet, Rao, Van~Beek, Nguyen, Garello, Kateel,
  Jayakumar, Costa, Cai, Yasin, Crotti, and Kar]{couet2021beol}
S.~Couet, S.~Rao, S.~Van~Beek, V.D. Nguyen, K.~Garello, V.~Kateel,
  G.~Jayakumar, J.D. Costa, K.~Cai, F.~Yasin, D.~Crotti, and G.S. Kar.
\newblock Beol compatible high retention perpendicular sot-mram device for sram
  replacement and machine learning.
\newblock In \emph{2021 Symposium on VLSI Technology}, pages 1--2, 2021.

\bibitem[Sakhare et~al.(2018)Sakhare, Perumkunnil, Bao, Rao, Kim, Crotti,
  Yasin, Couet, Swerts, Kundu, Yakimets, Baert, Oh, Spessot, Mocuta, Kar, and
  Furnemont]{sakhare2018enablement}
S.~Sakhare, M.~Perumkunnil, T.~Huynh Bao, S.~Rao, W.~Kim, D.~Crotti, F.~Yasin,
  S.~Couet, J.~Swerts, S.~Kundu, D.~Yakimets, R.~Baert, HR. Oh, A.~Spessot,
  A.~Mocuta, G.~Sankar Kar, and A.~Furnemont.
\newblock {Enablement of STT-MRAM as last level cache for the high performance
  computing domain at the 5nm node}.
\newblock \emph{2018 IEEE International Electron Devices Meeting (IEDM)},
  00:\penalty0 18.3.1--18.3.4, 2018.
\newblock \doi{10.1109/iedm.2018.8614637}.

\bibitem[Gupta et~al.(2020)Gupta, Perumkunnil, Garello, Rao, Yasin, Kar, and
  Furnémont]{mohit}
M.~Gupta, M.~Perumkunnil, K.~Garello, S.~Rao, F.~Yasin, G.S. Kar, and
  A.~Furnémont.
\newblock {High-density SOT-MRAM technology and design specifications for the
  embedded domain at 5nm node}.
\newblock \emph{2020 IEEE International Electron Devices Meeting (IEDM)},
  00:\penalty0 24.5.1--24.5.4, 2020.
\newblock \doi{10.1109/iedm13553.2020.9372068}.

\bibitem[Amar{\'u} et~al.(2015)Amar{\'u}, Gaillardon, and
  De~Micheli]{Amaru15benchmark}
Luca Amar{\'u}, Pierre-Emmanuel Gaillardon, and Giovanni De~Micheli.
\newblock {The EPFL combinational benchmark suite}.
\newblock In \emph{Proceedings of International Workshop on Logic and Synthesis
  (IWLS)}, 2015.

\bibitem[Raghavan et~al.(2015)Raghavan, Bardon, Jang, Schuddinck, Yakimets,
  Ryckaert, Mercha, Horiguchi, Collaert, Mocuta, Mocuta, Tokei, Verkest, Thean,
  and Steegen]{raghavan2015holisitic}
P.~Raghavan, M.~Garcia Bardon, D.~Jang, P.~Schuddinck, D.~Yakimets,
  J.~Ryckaert, A.~Mercha, N.~Horiguchi, N.~Collaert, A.~Mocuta, D.~Mocuta,
  Z.~Tokei, D.~Verkest, A.~Thean, and A.~Steegen.
\newblock {Holisitic Device Exploration for 7nm node}.
\newblock \emph{2015 IEEE Custom Integrated Circuits Conference (CICC)}, pages
  1--5, 2015.
\newblock \doi{10.1109/cicc.2015.7338377}.

\bibitem[Yang et~al.(2015)Yang, Ryu, and Parkin]{dw_velocity}
See-Hun Yang, Kwang-Su Ryu, and Stuart Parkin.
\newblock {Domain-wall velocities of up to 750 m s-1 driven by exchange
  coupling torque in synthetic antiferromagnets}.
\newblock \emph{Nature Nanotechnology}, 10\penalty0 (3):\penalty0 221--226,
  2015.
\newblock ISSN 1748-3387.
\newblock \doi{10.1038/nnano.2014.324}.

\bibitem[Bläsing et~al.(2020)Bläsing, Khan, Filippou, Garg, Hameed,
  Castrillon, and Parkin]{RacetrackMemory}
Robin Bläsing, Asif~Ali Khan, Panagiotis~Ch. Filippou, Chirag Garg, Fazal
  Hameed, Jeronimo Castrillon, and Stuart S.~P. Parkin.
\newblock {Magnetic Racetrack Memory: From Physics to the Cusp of Applications
  Within a Decade}.
\newblock \emph{Proceedings of the IEEE}, 108\penalty0 (8):\penalty0
  1303--1321, 2020.
\newblock ISSN 0018-9219.
\newblock \doi{10.1109/jproc.2020.2975719}.

\bibitem[Nikonov et~al.(2015)Nikonov, Manipatruni, and Young]{exchangedw}
Dmitri~E Nikonov, Sasikanth Manipatruni, and Ian~A Young.
\newblock {Cascade-able spin torque logic gates with input–output isolation}.
\newblock \emph{Physica Scripta}, 90\penalty0 (7):\penalty0 074047, 2015.
\newblock ISSN 0031-8949.
\newblock \doi{10.1088/0031-8949/90/7/074047}.

\bibitem[Vaysset et~al.(2018)Vaysset, Zografos, Manfrini, Mocuta, and
  Radu]{Vaysset}
Adrien Vaysset, Odysseas Zografos, Mauricio Manfrini, Dan Mocuta, and
  Iuliana~P. Radu.
\newblock {Wide operating window spin-torque majority gate towards large-scale
  integration of logic circuits}.
\newblock \emph{AIP Advances}, 8\penalty0 (5):\penalty0 055920, 2018.
\newblock \doi{10.1063/1.5007758}.

\bibitem[Prasad et~al.(2020)Prasad, Huang, Chopdekar, Chen, Steffes, Das, Li,
  Yang, Lin, Gosavi, Nikonov, Qiu, Martin, Huey, Young, Íñiguez, Manipatruni,
  and Ramesh]{prasad}
Bhagwati Prasad, Yen‐Lin Huang, Rajesh~V. Chopdekar, Zuhuang Chen, James
  Steffes, Sujit Das, Qian Li, Mengmeng Yang, Chia‐Ching Lin, Tanay Gosavi,
  Dmitri~E. Nikonov, Zi~Qiang Qiu, Lane~W. Martin, Bryan~D Huey, Ian Young,
  Jorge Íñiguez, Sasikanth Manipatruni, and Ramamoorthy Ramesh.
\newblock {Ultralow Voltage Manipulation of Ferromagnetism}.
\newblock \emph{Advanced Materials}, 32\penalty0 (28):\penalty0 2001943, 2020.
\newblock ISSN 0935-9648.
\newblock \doi{10.1002/adma.202001943}.

\bibitem[Zourafos et~al.(2017)Zourafos, Meester, Testa, Soekent, Gaillardon,
  De-Micheli, Amarù, Raghavan, Catthoor, and Lauwcreins]{pipe}
O.~Zourafos, A.~De Meester, E.~Testa, M.~Soekent, P.-E. Gaillardon,
  G.~De-Micheli, L.~Amarù, P.~Raghavan, F.~Catthoor, and R.~Lauwcreins.
\newblock {Wave Pipelining for Majority-Based Beyond-CMOS Technologies}.
\newblock \emph{Design, Automation \& Test in Europe Conference \& Exhibition
  (DATE), 2017}, pages 1306--1311, 2017.
\newblock \doi{10.23919/date.2017.7927195}.

\end{thebibliography}






\end{document}